\documentstyle[12pt,aaspp4]{article}
 
\lefthead{Schneider et al.}
\righthead{HI Mass Function of Galaxies}
 
\begin{document}
 
\title{ A Deep Survey of HI--Selected Galaxies: The HI Mass Function }

\author{Stephen E. Schneider, John G. Spitzak\altaffilmark{1},
and Jessica L. Rosenberg}
\affil{Five College Astronomy Program, Department of Physics and Astronomy,
University of Massachusetts, Amherst, MA 01003}
\author{{\it To appear in Astrophysical Journal Letters, November 1, 1998}}
 
\altaffiltext{1}{present address: Sea Beam Instruments, Boston, MA}
 
\begin{abstract}

We analyze the distribution of galaxy HI masses detected in
a large, deep HI survey conducted at the Arecibo observatory, and we
find possible evidence of a faint-end steepening of the mass function
similar to what has been found optically.
This is the first HI survey with enough dynamic range to see
this steepening; the results of an earlier survey are found to be
consistent when the detection statistics are re-examined.
We demonstrate a technique for testing and correcting source
count completeness in HI surveys based on the ${\cal V}/{\cal V}_{max}$ test
and the large scale structure in the regions surveyed.

\end{abstract}
 
\keywords{galaxies: luminosity function, mass function
--- large-scale structure of universe --- radio lines: galaxies}
 
\section{Introduction}
 
Recent optical studies indicate that the faint end of the galaxy luminosity
function may grow significantly steeper than the power-law slope
seen at intermediate ranges (Marzke, Huchra, \& Geller 1994a;
Driver \& Phillipps 1996; Loveday 1997). The rise appears to be present
in both cluster and field galaxies, and is particularly strong among
Magellanic spirals and irregulars (Marzke et al.~1994b).
This is intriguing because if the faint-end slope is steep enough, 
a significant fraction of baryonic matter may be bound to small galaxies.

Since optical counts of low-luminosity sources may have subtle selection
effects, several groups have taken an independent approach,
making ``blind'' searches for extragalactic 21 cm HI emission
(see Spitzak \& Schneider 1998, Paper I, and references therein).
HI observations are advantageous for detecting low-luminosity
systems because starlight is not needed to power 21 cm line emission,
and HI is generally abundant in field dwarfs.

Our HI survey in the ``Arecibo Slice''\footnote{The Arecibo Observatory is
part of the National Astronomy and Ionosphere Center, which is operated by
Cornell University under cooperative agreement with the National Science
Foundation.}
(Paper I) detected 75 HI sources and is the first blind HI survey to
detect sources over a range of masses comparable to the range of optical
luminosities in samples in which the faint-end steepening is seen.
One other published survey, the Arecibo HI Strip Survey or
``AHISS'' (Sorar 1994; Zwaan et al.~1997, hereafter ZBSS), samples a
comparable area of the sky, however it was carried out and analyzed in a
different fashion. In our Arecibo Slice,
closely spaced, pointed observations were made so that all sources were
detected in the telescope's main beam, and the effective
sensitivity over the survey area was relatively uniform.
In the AHISS, driftscans were used and sources were detected
all over the main beam and sidelobes of the telescope.

The most important information needed for converting survey detections into
a mass function is a thorough understanding of the survey's completeness.
Both the Arecibo Slice and AHISS have now been followed up by confirmation
observations that give accurate fluxes for the sources,
but to determine the completeness we must understand
the sensitivity to sources in their original detection scans.
For this reason the sidelobe detections of the AHISS are problematic.
The Arecibo sidelobes had asymmetries and temperature dependencies that
made their sensitivity uncertain.
Moreover, the original detection fluxes of AHISS sidelobe sources were
not saved (Sorar 1998, private communication), so we
cannot apply completeness tests after the fact.  For these
reasons, we will consider only the 45 main-beam AHISS sources in
the remainder of this paper.

The problem with sidelobe source sensitivity is also the main reason why an
analysis of the original results from the AHISS by Schneider (1997) gave
different results from ZBSS, who attributed the difference
to using the original Arecibo detection fluxes versus their new VLA
measurements. This is unlikely to explain the difference since gain
variations due to the uncertain positions should have introduced a
shift and a scatter
in the results that were smaller on average than the bin size used in
the mass function.  Instead, we believe the difference arises from
ZBSS's inclusion of low-sensitivity sidelobe detections---17 of 18
of these had high HI masses $M_{HI}>10^9M_\odot$. (Note that
we use Galactocentric velocities and assume $H_0=75$ km s$^{-1}$ Mpc$^{-1}$.)
Including these sources could easily bias the apparent slope of the HI
mass function because of the uncertainties in sidelobe behavior.

Rather than entering into further debate about details of observational
technique, we propose to demonstrate and apply a technique for directly
determining the completeness of HI surveys.  We show how to use an
estimate of the density variations due to large scale structure (LSS)
in the area of a survey to test and correct for survey sensitivity.
Based on the available data, we show that the Arecibo HI surveys are
consistent with a steepening at the faint end of the HI mass function,
which is quite similar to the results of Loveday (1997) for
optically-selected field galaxies.

We first examine the LSS in the regions of the two HI surveys in \S2.
Using the LSS information, we determine
the behavior of the sensitivity limits of the HI surveys in \S3. Then in
\S4, we derive the mass function for the Arecibo surveys.
We conclude with a summary and discussion of the results in \S5.

\section{Large Scale Structure in the Survey Regions}

Using optical redshifts we can make LSS density estimates as a function
of redshift over the areas surveyed for HI. These density estimates
should be appropriate for our purposes since HI sources qualitatively trace
the same structure as optically identified galaxies (Paper I).
We use galaxies from the ``RC3'' (de Vaucouleurs et al. 1991) that are
within $\delta=\pm10^\circ$ of the narrow declination bands examined in
each HI survey.
The RC3 sources with redshifts in these regions appear to be 
complete to a limiting magnitude of $m_{lim}\leq14.5$ based on their number
counts increasing like $10^{0.6m}$.
Since the AHISS strips cross through the Galactic plane the RC3
coverage of them is incomplete, but we shall show later that
the results are consistent with the Arecibo Slice and
the effect of LSS is small in any case.

For a galaxy with apparent magnitude $m_i$, the maximum redshift
at which it would remain brighter than $m_{lim}$ is
$z_{max,i}\equiv z_i 10^{(m_{lim}-m_i)/5}$,
where we assume the redshift $z_i$ is directly
proportional to distance.  The number of
galaxies we expect to observe with $z<z_i$ but $z_{max}>z_i$ is:
\begin{equation}
 N_i = {3\phi(M_i)\over z_i^3} \bar\rho(z_i)\;,
\end{equation}
where $\bar\rho(z_i)\equiv\int_0^{z_i}\rho(z)z^2\,dz$ is the mean
relative density out to redshift $z_i$. The quantity $\phi(M_i)$
represents an integral of the luminosity function,
but since it divides out in the end, its form need not be known.

\begin{figure}[tb]
\plotone{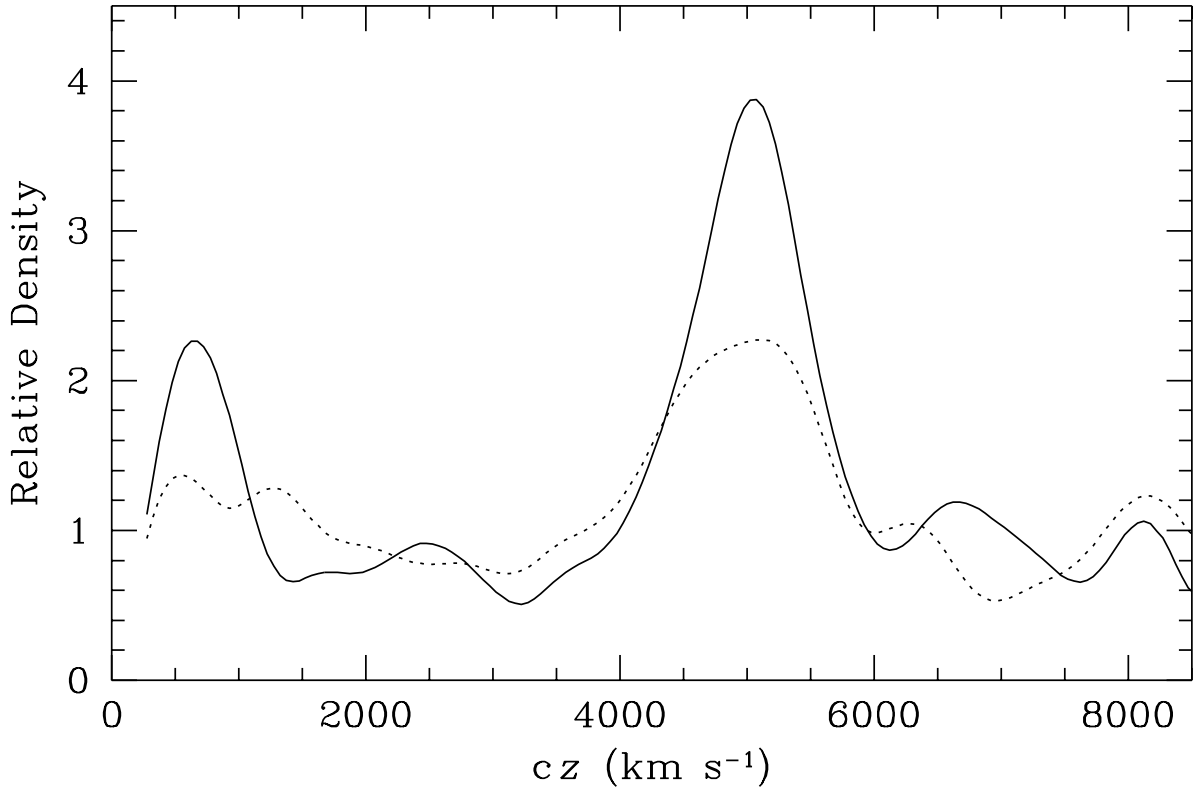}
\caption{
Estimated density of galaxies as a function of redshift in the regions
surrounding the HI surveys. The solid line is for the Arecibo Slice
(Paper I), and the dashed line is for the AHISS.
}
\label{LSSdensity}
\end{figure}

We can write a recursive relation between the mean density for the
galaxies out to $z_i$ and out to the galaxy with the next smaller
redshift $z_{i-1}$:
\begin{equation}
\bar\rho(z_{i-1})=\bar\rho(z_i){(N_i-1)/z_i^3\over N_i/z_{i-1}^3} \;.
\end{equation}
 We normalize $\bar\rho=1$ averaged over the full redshift range of the survey.
(This normalization may affect the overall scaling of the mass function,
but not its shape.)
This gives us the mean relative density for all smaller redshifts
which we differentiate to find the local relative
density $\rho(z)$ shown in Fig.~\ref{LSSdensity}. 
(To produce this plot the data were binned and smoothed 
with a Gaussian of 500 km s$^{-1}$ FWHM.)
Both surveys show an excess of galaxies around
$v=5000$ km s$^{-1}$ caused by the Pisces-Perseus supercluster.

The density of galaxies also rises at small redshifts in the Arecibo Slice
due to the local supercluster.
This nearby rise might exaggerate the counts of low mass sources since they
are detectable only out to small velocities. However, we shall show in
\S 5 that the magnitude of the effect is small. A more useful aspect
of our LSS determination is that it permits us to test 
the surveys' completeness.

\section{HI Survey Completeness }

To determine the number density of sources of a given HI mass,
we must accurately establish the volume within which a survey is
sensitive to them. This is complicated by several factors: (1) the same galaxy
observed at different inclinations will have 
different line widths; (2) blind HI surveys detect objects with
sensitivities that depend on distance from beam center and frequency;
and (3) efforts to reject man-made interference and to subtract
instrumental ``baselines'' can artificially suppress real signals.
Since these  HI surveys include only confirmed sources, we do not need
to worry about the opposite problem of including unreliable sources,
but we need some means of establishing the surveys' completeness.

For a galaxy with a total signal $S$ (integrating the HI flux density
over the 21 cm line width $w$) the peak signal-to-noise ratio $S/N$ is
achieved when the spectrum is smoothed to a velocity resolution equal to $w$.
In this ideal case, $S/N \propto w^{-1/2}$
(see, for example, Schneider 1997; ZBSS).
In Paper I we found our minimum detected fluxes were at a signal-to-noise
$S/N\gtrsim5$ based on this description.

The noise level in the AHISS spectra varied substantially in different
regions depending on the total number of drift-scan spectra, which
were collected once per day and then averaged.
When we use the noise value for each source 
(based on the number of observations averaged in its vicinity according
to Sorar 1994) we find a lower limit of $S/N\gtrsim7$.
This higher limit probably reflects the fact that Sorar (1994) confirmed
whether sources were detected ``by inspecting the data for each day
separately.'' Sources close to his quoted 5$\sigma$ search limit would
be only 1--2$\sigma$ in a single day's data and very difficult to confirm.

A direct test of whether a survey is complete to its quoted 
sensitivity limit is to compute
${\cal V}/{\cal V}_{max} = (z/z_{max})^3$. This should average to $1/2$ if
$z_{max}$ is correctly determined (Schmidt 1968).
In this test we set $z_{max}$ to the smaller of the bandpass limit and
the highest detectable redshift based on the claimed detection limit.
We find ${\cal V}/{\cal V}_{max}$ averages to 0.39 for the Arecibo Slice
using our quoted 5$\sigma$ limit. We find an even lower average, 0.34,
for the AHISS sources using their quoted 5$\sigma$ limit, or 0.41 using our
estimated 7$\sigma$ limit.

\begin{figure}[tbp]
\plotone{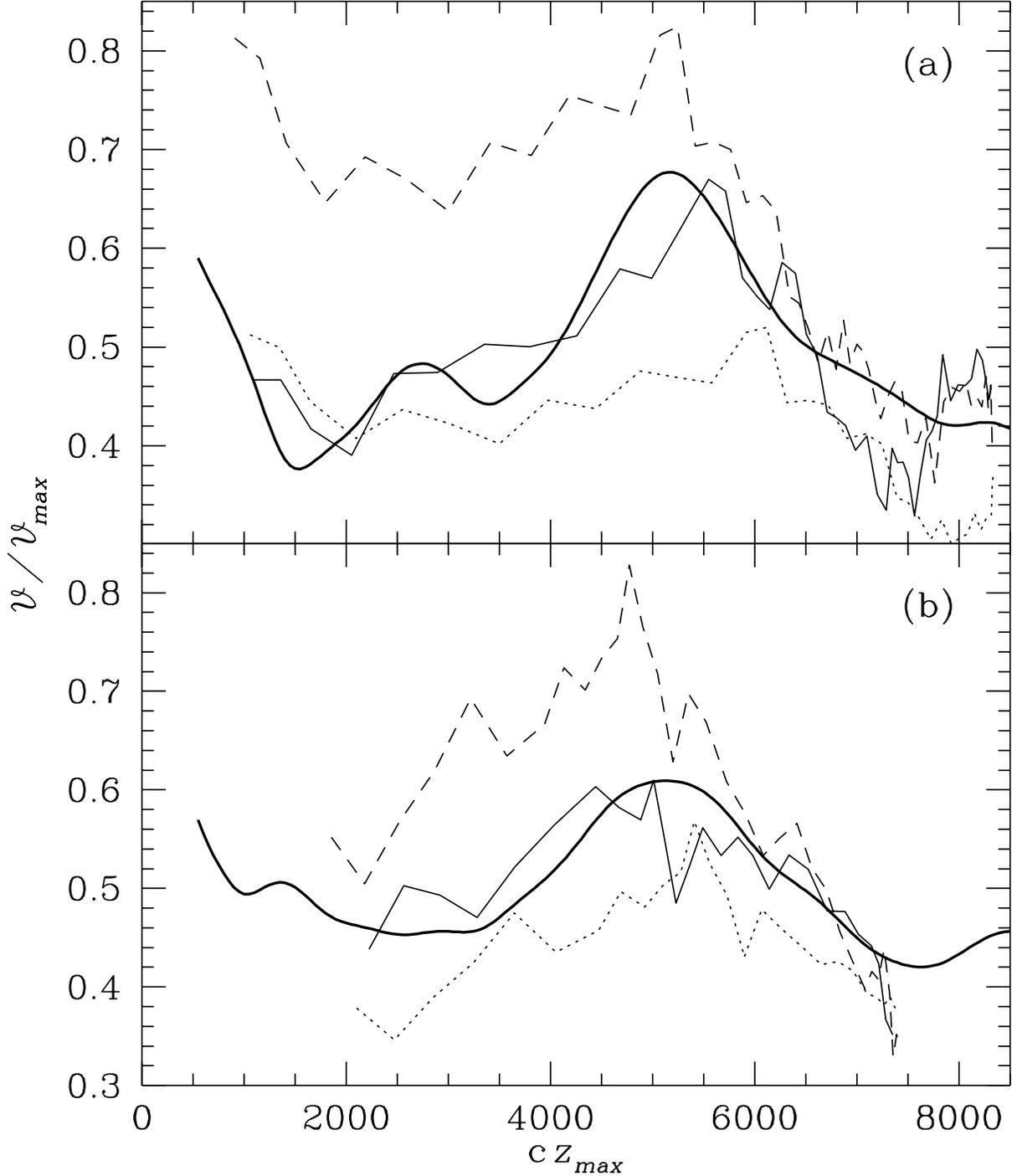}
\caption{
Expected and observed values of ${\cal V}/{\cal V}_{max}$ as a function
of $z_{max}$ for (a) the Arecibo Slice, and (b) the AHISS.
The heavy solid line shows the expected distribution
derived from the density structure in Fig.~\protect{\ref{LSSdensity}}.
The dotted and dashed lines show the distribution determined from
the HI data assuming 5 and 7--$\sigma$ completeness limits
(7 and 10--$\sigma$ for the AHISS). The thin solid line shows
the results for an empirical model where the detection depends on the
line width to the 0.75 power (see text).
}
\label{VVmax}
\end{figure}

To determine whether LSS could affect ${\cal V}/{\cal V}_{max}$,
we examine its dependence on redshift. The heavy
line in Fig.~\ref{VVmax} shows the expected behavior as a function of
redshift for the density distribution derived in \S2. (At each $z_{max}$
we consider all sources $i$ with $z_i<z_{max}<z_{max,i}$.)
The dotted lines show the observed behavior of ${\cal V}/{\cal V}_{max}$
based on the suggested detection limits of the two HI surveys. The values
are systematically below $1/2$ even at redshifts where LSS should make
them high.

This should not be surprising, since a {\it detection} limit is normally
lower than a {\it completeness} limit.  We can force the average value of
${\cal V}/{\cal V}_{max}$ to be $\approx1/2$ by assuming the $S/N$ must
be 40\% higher for the completeness limit (dashed lines), but the result is
systematically too high at low redshifts.
A better explanation is that the completeness limit does not actually depend
on a source's line width like $w^{0.5}$ as assumed above.
We noted indications of this in Paper I---narrow-line sources were
detected to lower $S/N$.
We have experimented with sensitivities that depend on $w^x$, defining a
new ``signal-to-noise'' ratio:
\begin{equation}
S/N_{new}=
 {(300\hbox{km s}^{-1}/w_{res})^{x-0.5}\int S\,dv
 \over (w/w_{res})^{x}\times \hbox{rms}\times w_{res}}
\label{SNeqn}
\end{equation}
where rms is the noise measured at the spectral resolution
$w_{res}$, $\int S\,dv$ is the integrated flux in the line, and
the result is normalized to match $S/N$ at $w=300$ km s$^{-1}$.

A good empirical fit for the completeness limit is found with $x=0.75$,
which suggests that a minimum flux density also plays a role in detection.
For $x=0.75$ and $S/N_{new}=7$ ($S/N_{new}=10$ for AHISS detections),
we get the thin solid-line curves in Fig.~\ref{VVmax}, which
match the expected ${\cal V}/{\cal V}_{max}$ behavior quite well.
By inverting eqn.~\ref{SNeqn}, these values also define the 
``completeness flux'' for a source, $(\int S\,dv)_{comp}$,
for which the survey is on average complete.

\section{The HI Mass Function}

Equipped with knowledge of LSS and survey completeness, we can 
derive the HI mass function. We do this by means of the ${\cal V}_{tot}$
method. ZBSS refer to this as the $\Sigma 1/{\cal V}_{max}$
method, but we want to avoid confusion with ${\cal V}_{max}$ as defined
in \S3.
${\cal V}_{tot}$ describes the total ``completeness volume'' in which a
source should on average have been detected over the entire
survey area, taking into account offsets from beam center, etc.;
${\cal V}_{max}$ describes the maximum distance a source
should have been detected in its actual detection spectrum.
The completeness volume of each source implies an overall density
of sources:
\begin{equation}
\Phi(M_1<M<M_2)=\sum_{M_1<M_i<M_2}{1/{\cal V}_{tot}(M_i)}
\end{equation}
for sources in the mass range $M_1$--$M_2$.
This method depends critically on an accurate determination of the
completeness volume, which we discuss first, and is also influenced
by LSS, which we address subsequently.

We empirically determined the decline in sensitivity to sources offset from
the beam center by using the ratio of detection fluxes to true fluxes
(found from follow-up observations). Because of the hexagonal
sampling pattern in the Arecibo Slice, sources were never farther than
about 2.3$'$ from beam center. We limit our analysis of AHISS
sources to those within 3$'$ of beam center.

The frequency dependence of Arecibo's line feeds changes the gain
by up to a factor of $\sim2$ over the observed redshift ranges.
The AHISS $\delta=23^\circ$ strip presents a problem here.  The data were
collected with the feed tuned to different frequencies on different days,
making the sensitivity up to $2\times$ larger or smaller at different
points in the bandpass, and the detections were carried out on subsets
of the data with different frequency dependences. Too few details of
this were recorded to allow us to reconstruct the exact detection
characteristics, so we treat the $\delta=23^\circ$ strip like the
$\delta=14^\circ$ strip, for which the frequency dependence remained fixed.

We determine ${\cal V}_{tot}$ for each source in the Arecibo Slice and 
AHISS by numerically integrating the search volume over all possible
positions within the survey boundaries. We include the frequency dependence
across the bandpass to determine at which redshifts a source's observed
flux should be detectable. Finally, the differing AHISS $rms$ noise levels
were also included in the calculation of the total completeness volume.

Some of the lowest flux (not necessarily low mass) sources are
fainter than the completeness flux, $(\int S\,dv)_{comp}$,
estimated above. However, this does not affect the shape of the mass
function at the faint end since ${\cal V}_{tot}$ changes by the same
factor for all but the highest mass sources ($\gtrsim10^9M\sun$).
We checked this conclusion by using Monte Carlo
simulations of sources following a Schechter function ($\alpha=-1.1$ to
$-1.5$), selecting sources statistically to imitate the declining
completeness seen in the HI surveys. We found that application of the
${\cal V}/{\cal V}_{max}$ test and ${\cal V}_{tot}$ method recovered the
input density and slope.

Another concern in determining the mass function is the effect of large
scale structure. ZBSS have shown that the ${\cal V}_{tot}$ method is
fairly insensitive to sinusoidal variations in LSS, but it is important
to determine whether a local overdensity could have a
significant effect on the source counts at the low-mass end.

Our test of the possible significance of LSS is to use the optically-based
density structure determined in \S2. Each source's value of
${\cal V}_{tot}$ is corrected to account for the probability of finding a
source within its completeness volume. This is done by dividing
${\cal V}_{tot}$ by the mean density out to the source's maximum
detectable redshift: $\bar\rho(z_{max})$.  Since this
depends on beam offset and the other
sensitivity factors, we calculate the corrected volume
in our numerical integrations of the completeness volume described above.

\begin{figure}[tbp]
\plotone{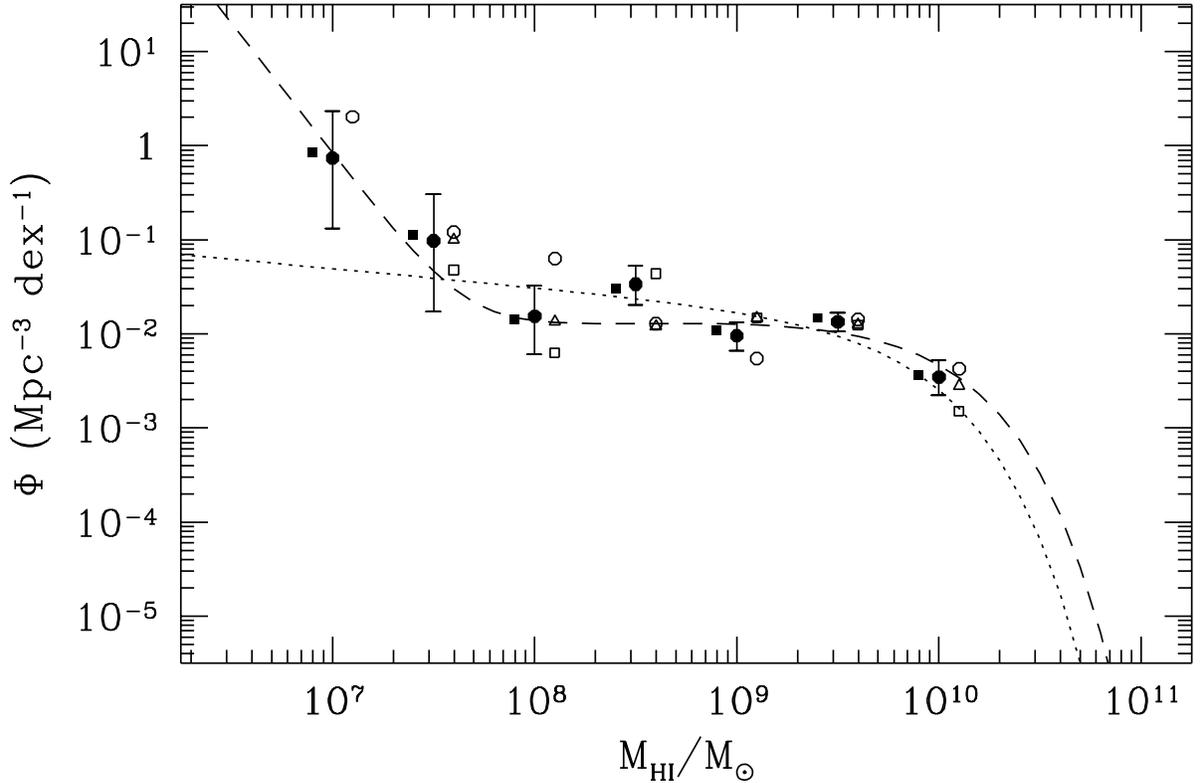}
\caption{
The HI mass function determined from Arecibo surveys of HI sources.
Solid circles with 95\% confidence limits show the joint mass function
of the Arecibo Slice and AHISS. Several subsets and alternative
reductions of the data are also shown without error bars, and are
offset right and left of the joint mass function points for clarity.
Open circles show the mass function
of the Arecibo Slice alone; open squares, AHISS; open triangles, AHISS
restricted to the $\delta=14^\circ$ strip. Solid squares show
the mass function derived when no corrections are made for large
scale structure.
The dotted curve shows a Schechter function with
$\alpha=-1.2$ suggested by ZBSS, and the dashed curve
is the functional form of the faint-end turn-up found by Loveday (1997)
for optical sources.
}
\label{HImassfunc}
\end{figure}

The joint mass function for the Arecibo Slice and AHISS
is shown by solid circles with Poisson 95\% confidence limits
(Gehrels 1986) in Fig.~\ref{HImassfunc}.
To calculate the joint
mass function requires that we determine each source's completeness
volume in the other survey. This is straightforward since line widths
and source fluxes were measured similarly in the two surveys, giving
us a sample of 120 HI sources.

The LSS turns out to have only a minor effect on the resulting mass
function. Solid squares show the mass function we would have determined
directly from the ${\cal V}_{tot}$ method with no LSS corrections.
(Note that the points are offset to the left for clarity.)
There is a small adjustment at low-masses,
but the local overdensity would have to be many times larger for this to
have generated the higher value we see.

The results from the Arecibo Slice alone are shown by open circles in the
figure; from AHISS, by open squares. We also show results for the
$\delta=14^\circ$ strip of AHISS (open triangles) alone, which did not
suffer from the problems with the $\delta=23^\circ$ strip described earlier.
The $\delta=14^\circ$ results generally agree better with the Arecibo
Slice results.

\section{Summary and Discussion}

The present combination of the Arecibo Slice and Arecibo HI Strip
Survey yields the largest and deepest sample of HI-selected sources
studied to date. The HI mass function derived from these surveys is
suggestive of a turn-up at the low-mass end of the HI mass function
similar to that seen at the faint end of the optical luminosity function.
In Fig.~\ref{HImassfunc} we show the empirical luminosity function found
by Loveday (1997) scaled to match the turnover at high HI masses. The
turn-up we see at low HI masses is at about the same relative
location Loveday found optically.
The rise at the faint end of the HI mass function appears to be in
good agreement with the optical results of Marzke et al.~(1994b), who
found that the
rise results primarily from late-type galaxies.

ZBSS have argued that the HI mass function is consistent with a
relatively shallow-sloped rise like the $\alpha=-1.2$ Schechter function 
shown by a dotted curve in Fig.~\ref{HImassfunc}. We find an
apparent rise in the lowest $10^7 M\sun$ bin, in which two HI sources were
detected in the Arecibo Slice. The AHISS detected no sources with such
low masses, but we would have expected only 1--2 sources ($<1$ source
in the $\delta=14^\circ$ strip), based on the relative volume
sensitivities. Obviously these are small number statistics, but the
95\% Poisson confidence limit excludes the Schechter fit proposed by ZBSS.

The differences we find in this paper result from the greater success of the
Arecibo Slice project in detecting low-mass HI sources, the rejection
of sidelobe sources from the AHISS, and the 
completeness tests we have applied. We believe that many
previous surveys have overestimated their sensitivity and hence their
completeness volumes (${\cal V}_{tot}$), and therefore underestimated
the counts of low mass sources.  The rise at the low-mass
end of the HI mass function is still a relatively weak statistical result,
but we have shown that it cannot be attributed
to two potential observational issues. (1) Narrow-line sources are easier to
detect than wide-line sources at the same statistical $S/N$. We have
determined that this is described by an effective
``$S/N$''$\propto w^{0.75}$ and adjust our results accordingly.
(2) The overdensity of sources in the Local Supercluster can also
exaggerate low-mass counts, but we show this effect to be minimal, at
least if HI sources are no more clustered than optically-selected sources.

The volume of space so-far surveyed at high enough sensitivity to
detect low HI-mass sources is still small, so large new surveys
currently planned or ongoing should definitively establish the shape
of HI mass function. However, the results will remain only as accurate
as the determination of the surveys' actual completeness limits.
We have demonstrated in this paper how to use information about the LSS
and a ${\cal V}/{\cal V}_{max}$ test to directly determine an appropriate 
sensitivity limit, and we recommend that other HI surveys test their
completeness similarly.

\end{document}